--------------------------------------------------------------------------------------
\documentstyle[eqsecnum,aps]{revtex}
\begin{document}

\title{Algebraic approach to vibrational spectra 
of tetrahedral molecules: \\
a case study of silicon tetrafluoride}

\author{Xi-Wen Hou~$^{a,b}$, Shi-Hai Dong~$^a$, Mi Xie~$^c$, and 
Zhong-Qi Ma~$^a$}

\address{ ~$^a$ Institute of High Energy Physics, 
P.O. Box 918(4), Beijing 100039,The People's Republic of China\\
~$^b$ Department of Physics, University of Three Gorges, 
Yichang 443000, The People's Republic of China \\
~$^c$ Department of Physics, Tianjin Normal University, 
Tianjin 300074, The People's Republic of China }

\maketitle

\vspace{5mm}
\begin{abstract}
Both the stretch and bend vibrational spectrum 
and the intensity of infrared transitions in a tetrahedral 
molecule are studied in a U(2) algebraic model, 
where the spurious states in the model Hamiltonian 
and the wavefunctions are exactly removed. As an example,
we apply the model to silicon tetrafluoride SiF$_4$.
\end{abstract}

\vspace{5mm}
\section{INTRODUCTION}

In recent years, algebraic methods [1,2] have been developed to be
an effective theory for descriptions of vibrations, rotations, 
and rotation-vibration interactions in polyatomic molecules [3,4]. 
U(4) and U(2) algebraic model have mostly been used so far in the 
analysis of experimental data. U(4) model [5,6] takes the rotation and
the vibration into account simultaneously but becomes quite complex
when the number of atoms in the molecules becomes larger than 4, while
U(2) model is particularly well suited for dealing with the stretching
vibrations of polyatomic molecules such as the octahedral 
and benzene-like systems [7,8]. Those two models are still extensively 
used for small molecules [9,10]. Recently,
an active step in the development of algebraic models has been taken 
by Iachello and Oss to incorporate the bending modes 
into the models for benzene [11] and acetylene [12]. 
Extending U(2) algebraic model [3] to describe both the stretching 
and the bending vibrations in X$_3$ [13] and X$_4$ molecules [14], 
Frank and his co-workers have introduced a symmetry-adapted 
algebraic model, and established an explicit connection between 
algebraic and configuration space calculations. Developing the local 
mode model [15], we have recently proposed an algebraic model of boson 
realization [16-18] for the complete vibrations by 
taking full advantages of the discrete symmetries of molecular systems.
It is worth mentioning that U(k-1) and U(n) algebraic approach 
have been introduced for the k=3m-3 rotational and vibrational degrees 
of freedom of m-atomic molecules [19], and for n-1 stretching vibrational 
degrees of freedom of polyatomic molecules [20], respectively. However,
those two approaches are less feasible than U(4) and U(2) model in 
application.

In order to study vibrational modes of medium-size molecules, 
one usually chooses a tetrahedral molecule as one of good samples.
Its excited stretching vibrational states were 
explained in the local mode model [21]. 
In our previous papers, the vibrational spectrum of methane was 
analyzed in terms of bosonic operators [16] and q-deformed 
harmonic oscillators [23], however its infrared intensity was not 
taken into account. Recently new algebraic SU(2) model [24]
has been proposed for tetrahedral molecules. 
This model needs at least eight parameters for the calculation of
transition intensities. The lack of experimental data did not allow
such a fit for the methane molecule, and 
interactions between the stretch and the bend were neglected 
in the calculation of energy levels. Leroy $et$ $al.$ 
have up to date presented an algebraic model [22] only for highly 
excited stretching modes and infrared intensities in this system. 
Although a precise infrared transition model 
has been constructed in algebraic scheme, it will be complex
when the bending vibrations are considered.

In this letter, both stretching and bending vibrations and intensities 
of infrared transition of all active modes in a tetrahedral molecule  
are studied in a U(2) algebraic model based on our new 
method for eliminating the spurious states and spurious 
components in Hamiltonian. Our model transition operator contains only 
five parameters. One of its applications is presented  
to silicon tetrafluoride SiF$_4$. Calculated results demonstrate
that our method for dealing with spurious
states are also effective for U(2) algebraic model, which open the 
possibility to use this model for analyzing the complete vibrations 
in large molecules where spurious states exist.  
In Sec. II, considering the symmetry of a tetrahedral molecule,
we set up the model Hamiltonian, in which the interactions related to
spurious states are exactly removed. Its infrared intensity is studied in
Sec. III. Application to SiF$_4$ is given in Sec. IV, where
symmetrized bases are used to simplify calculations.
Concluding remarks are made in Sec. V.
 
\section{HAMILTONIAN}

For a tetrahedral molecule XY$_4$, we introduce ten U(2)
algebras to describe ten interactions between atoms: 
U$_j$(2) ($1\leq j \leq 4$) for X-Y and U$_{\mu}$(2) 
($5\leq \mu \leq 10$) for Y-Y interactions. The assignments 
of the Cartesian coordinate system are the
same as in Ref. [16]. Each U$_{\alpha}$(2) ($1\leq \alpha \leq 10$) is
generated by the operators $\{ \hat{N}_{\alpha},\hat{J}_{+,\alpha} 
\hat{J}_{-,\alpha}, \hat{J}_{0,\alpha} \}$, 
satisfying the following commutation relations:  
$$\begin{array}{ll}
~[\hat{J}_{0,\alpha}, \hat{J}_{\pm,\beta}]=\pm \delta_{\alpha \beta}
\hat{J}_{\pm,\alpha},~~~
&[\hat{J}_{+,\alpha}, \hat{J}_{-,\beta}]=2 \delta_{\alpha \beta} 
\hat{J}_{0,\alpha},\\
~[\hat{N}_{\alpha}, \hat{J}_{0,\beta}]=0,  
~~&[\hat{N}_{\alpha}, \hat{J}_{\pm,\beta}]=0 .
\end{array}  $$

\noindent
where $\hat{N}_{\alpha}$ is related with the Casimir operator
of U(2):
$$2\hat{J}_{0,\alpha}^{2}+\hat{J}_{+,\alpha}\hat{J}_{-,\alpha}
+\hat{J}_{-,\alpha}\hat{J}_{+,\alpha}
=\hat{N}_{\alpha}(\hat{N}_{\alpha}/2+1). $$

\noindent
Denote by $v_{\alpha}$ the number of quanta in the $\alpha$th
bond. The local basis states for each bond are labeled by 
the eigenvalue $N_{\alpha}$ of $\hat{N}_{\alpha}$
and $v_{\alpha}$, and written as $|N_{\alpha},v_{\alpha}\rangle$. 
Their products provide the local bases: 
$$|N_1,v_1\rangle  |N_2,v_2\rangle \cdots 
|N_{10},v_{10}\rangle  \equiv  
|N_{\alpha},v_{\alpha}\rangle . $$

\noindent
Those $N_{j}$ of equivalent bonds are equal to each other:
$N_{j}= N_s$, and $N_{\mu}= N_b$, 
where and hereafter, the indexes $j$, $\mu$, and $\alpha$ run from 1 to 4, 
5 to 10, and 1 to 10, respectively, and footnotes $s$ and $b$ refer to
the X-Y couplings and the Y-Y couplings, respectively.

There are three kinds of O(2) invariant combinations
of those generators: 
\begin{equation}
\begin{array}{rl}
\hat{H}_{\alpha}^{M}&=~(\hat{J}_{+,\alpha}\hat{J}_{-,\alpha}  
+\hat{J}_{-,\alpha}, \hat{J}_{+,\alpha})/2~-~\hat{N}_{\alpha}/2, \\
\hat{H}_{\alpha,\beta}&=~2\hat{J}_{0,\alpha}\hat{J}_{0,\beta}
~-~ \hat{N}_{\alpha}\hat{N}_{\beta}/2,~~~ \alpha \neq \beta,  \\
\hat{V}_{\alpha,\beta}&=\hat{J}_{+,\alpha}\hat{J}_{-,\beta}~+~  
\hat{J}_{-,\alpha}\hat{J}_{+,\beta}, ~~~\alpha \neq \beta.
\end{array}
\end{equation}

\noindent
Their matrix elements in the local bases are given in Ref. [13].  
The operators $\hat{H}_{\alpha}^M$ correspond to the energy of the
$\alpha$th Morse oscillator. The operators $\hat{H}_{\alpha,\beta}$
describe anharmonic terms of the type $v_{\alpha}v_{\beta}$, while the 
operators $\hat{V}_{\alpha,\beta}$ describe interbond couplings which, in 
configuration space, are of the type ${\bf r}_{\alpha}{\bf r}_{\beta}$, 
where ${\bf r}_{\alpha}$ and ${\bf r}_{\beta}$ are the displacement
vectors of bonds $\alpha$ and $\beta$ from their equilibrium positions.
 
Considering that $\sum \hat{J}_{+,\mu}$ (or $\sum \hat{J}_{-,\mu}$) relates 
only to spurious states [23], we obtain the following $T_{d}$ invariant 
Hamiltonian in terms of those three kinds of operators:
\begin{equation}
\begin{array}{rl}
H&=~H_s~+~H_b~+~H_{sb},  \\
H_s&=~\displaystyle \lambda_{s1}~ \sum_{j=1}^{4}~ \hat{H}_j^M
~+~\lambda_{s2}~\displaystyle \sum_{i\neq j=1}^{4}~ \hat{H}_{i,j}
~+~\lambda_{s3}~\displaystyle \sum_{i\neq j=1}^{4}~ \hat{V}_{i,j},  \\
H_b&=~\displaystyle \lambda_{b1}~ \sum_{\mu=5}^{10}~ H_{\mu}^M
~+~\lambda_{b2}~\displaystyle \sum_{\nu-\mu \neq 3,\mu<\nu=6}^{10}~
\hat{H}_{\mu,\nu}    \\
&~~~~+~\lambda_{b3}~\displaystyle \sum_{\mu=5}^{7}~
\hat{H}_{\mu,\mu+3}   
~+~\lambda_{b4}~\displaystyle \sum_{\mu=5}^{7}~
\hat{V}_{\mu,\mu+3}, \\
H_{sb}&=~\lambda_{sb1}~ \left\{\displaystyle 
\sum_{\mu =5}^{7} \left(\hat{H}_{1,\mu}-\hat{H}_{1,\mu +3}\right)
~+~ \left(\hat{H}_{2,5}-\displaystyle \sum_{\mu =6}^{8} \hat{H}_{2,\mu}
+\hat{H}_{2,9}+\hat{H}_{2,10}\right)\right. \\
&~~~~+\left. \displaystyle \sum_{\mu =3}^{5}\left(\hat{H}_{3,2\mu}-
\hat{H}_{3,2\mu -1}\right)
~+\left(-\hat{H}_{4,5}-\hat{H}_{4,6}+
\displaystyle \sum_{\mu =7}^{9} \hat{H}_{4,\mu}
-\hat{H}_{4,10}\right)\right\}  \\
&~~~~+~\lambda_{sb2}~ \left\{\displaystyle 
\sum_{\mu =5}^{7} \left(\hat{V}_{1,\mu}-\hat{V}_{1,\mu +3}\right)
~+~ \left(\hat{V}_{2,5}-\displaystyle \sum_{\mu =6}^{8} \hat{V}_{2,\mu}
+\hat{V}_{2,9}+\hat{V}_{2,10}\right)\right.  \\
&~~~~+\left. \displaystyle \sum_{\mu =3}^{5}\left(\hat{V}_{3,2\mu}-
\hat{V}_{3,2\mu -1}\right)
~+\left(-\hat{V}_{4,5}-\hat{V}_{4,6}+
\displaystyle \sum_{\mu =7}^{9} \hat{V}_{4,\mu}
-\hat{V}_{4,10}\right)\right\},   
\end{array}  
\end{equation}

\noindent
where $H_s$, $H_b$, and $H_{sb}$ describe the stretching interaction, 
the bending one, and the interaction between the stretching and bending 
modes, respectively. The Hamiltonian preserves the quantum number 
$V=\sum v_{\alpha}$.

\section{INTENSITIES OF INFRARED TRANSITION}

The infrared active mode is $F_2$. The absolute absorption 
intensities from state $v'$ to $v$ are given by
\begin{equation}
\begin{array}{rl}
I_{vv'}&=~\nu_{vv'}P_{vv'} ,  \\
P_{vv'}&=~|\langle v|\hat{T}_x|v'\rangle|^{2}+
|\langle v|\hat{T}_y|v'\rangle |^{2}+
|\langle v|\hat{T}_z|v'\rangle |^{2},
\end{array} 
\end{equation}

\noindent
where $\nu_{vv'}$ is the frequency of the observed transition,
$\hat{T}_x$, $\hat{T}_y$, and $\hat{T}_z$ correspond to the three
components of the infrared transition operator $\hat{T}$, and
the state $|v\rangle$ denotes $|N_{\alpha}, v_{\alpha}\rangle$ 
for short. All other
constants are absorbed in the normalization of the operator $\hat{T}$.
The three components of $\hat{T}$ are
\begin{equation}
\begin{array}{rl}
\hat{T}_x&=~\zeta_s~(\hat{t}_1-\hat{t}_2+\hat{t}_3-\hat{t}_4)
            ~+~\zeta_b~(\hat{t}_6-\hat{t}_9)
~+~\zeta_{sb}~(\hat{t}_1+\hat{t}_2+\hat{t}_3+\hat{t}_4)(\hat{t}_6
        -\hat{t}_9), \\
\hat{T}_y&=~\zeta_s~(\hat{t}_1-\hat{t}_2-\hat{t}_3+\hat{t}_4)
            ~+~\zeta_b~(\hat{t}_7-\hat{t}_{10})
~+~\zeta_{sb}~(\hat{t}_1+\hat{t}_2+\hat{t}_3+\hat{t}_4)(\hat{t}_7
      -\hat{t}_{10}) , \\
\hat{T}_z&=~\zeta_s~(\hat{t}_1+\hat{t}_2-\hat{t}_3-\hat{t}_4)
            ~+~\zeta_b~(\hat{t}_5-\hat{t}_8)
~+~\zeta_{sb}~(\hat{t}_1+\hat{t}_2+\hat{t}_3+\hat{t}_4)(\hat{t}_5
     -\hat{t}_8),
\end{array} 
\end{equation}

\noindent
where $\zeta_s$, $\zeta_b$, and $\zeta_{sb}$ are parameters,
and $\hat{t}_{\alpha}$ is the local operator for the $\alpha$th bond.
The term with $\zeta_{sb}$ is one of the higher order contributions 
of $\hat{T}$, which is necessary for describing both the stretching and 
the bending active modes. The matrix elements of $\hat{t}_{\alpha}$ are 
taken to be [11]
\begin{equation}
\langle \hat{N}_{\alpha},v_{\alpha}|\hat{t}_{\alpha}
|\hat{N}_{\alpha},v'_{\alpha}\rangle~=~exp(-\eta_{\alpha}
|v_{\alpha}-v'_{\alpha}|).
\end{equation}

\noindent
Those $\eta_{j}$ for equivalent bonds are equal to each other: 
$\eta_{j}\equiv \eta_s$, and $\eta_{\mu}\equiv \eta_b$. 

The five parameters in the transition operator of Eqs. (3.2)-(3.3)
will be determined by fitting observed data. The calculated intensity
can be used to check assignments and in the study of intramolecular
energy relaxation in tetrahedral molecules.

\section{APPLICATION}

We now apply this model to study the complete vibrational spectrum and
the infrared intensity of SiF$_4$. To our knowledge, there are sixteen
observed vibrational energy levels and thirteen infrared intensity data 
for SiF$_4$ [25].

At first, we calculate the Hamiltonian matrix. The calculation for 
energy levels will be greatly simplified if the symmetrized bases 
are used. For the stretch and bend states of tetrahedral molecules, 
the symmetrized bases were given in Ref. [16], where the spurious 
states in the bases were eliminated. For stretching overtones of 
large molecules, the symmetrized bases have been recently presented 
by Chen {\it et al.} [26,27]. In those symmetrized bases the Hamiltonian 
matrix is a block one. According to Morse potentials for the stretch
and the bend vibrations in the boson-realization model [28] for this 
molecule, we take two boson numbers $N_s$ and $N_b$ to be 100 and 
15, respectively. The parameters in Hamiltonian are 
determined by fitting the observed data, and given in cm$^{-1}$
as follows:
$$\begin{array}{lllll}
\lambda_{s1}=-3.311, ~~~&\lambda_{s2}=-4.331, ~~~&\lambda_{s3}=-0.573,
~~~&\lambda_{b1}=97.982, ~~~&\lambda_{b2}=-1.197,\\ 
\lambda_{b3}=74.438,~~~&\lambda_{b4}=-4.173, 
~~~&\lambda_{sb1}=-0.145, ~~~&\lambda_{sb2}=0.199 .&
\end{array}  $$

\noindent
The experimental data and the calculated values are listed in 
Table I. The standard deviation in this fit is 1.188 cm$^{-1}$.
From those parameters we can calculate the other values. 
It is worth mentioning that McDowell {\it et al.} [25]
described the same energy levels by Dunham expansion with more 
parameters. This method does not provide explicitly wave
functions so that some physical properties such as transition
intensities are hard to be calculated.

\begin{center}

\fbox{Table I}

\end{center}

Then, we compute the infrared intensity. Due to the
Wigner-Eckart theorem, it is sufficient to calculate
only the $z$ component of the transition operator, $\hat{T_z}$,
in the symmetrized bases. Fitting the thirteen observed
infrared intensities, we determine the parameters in
the transition operator as follows:
$$ \eta_s=40.158, ~~~\eta_b=22.853,
~~~\zeta_s=3.614, ~~~\zeta_b=3.286, ~~~\zeta_{sb}=64.959.      $$ 

\noindent
In Table I we only list those calculated intensities to compare
with known observed data. The other calculated intensities and
energy levels can be obtained from us upon request.

One may observe in Table I that most of the calculated intensities 
are in good agreement with the experimental values, but
a few are not. Those differences may come from that the observed 
intensities were only approximately accurate [25], and that the other 
higher order contributions to the operator $\hat{T}$ were neglected.
It is worth pointing out that there are also differences in magnitude
between the calculated and the observed intensities for the stretching 
vibrations of octahedral molecules in U(2) model [29] because the same 
simple matrix elements Eqs.(3.3) were used.  
The more accurate experimental data are needed to improve the model. 

\section{CONCLUDING REMARKS}

We have used a U(2) algebraic model for studying the
stretching and bending vibrations and infrared intensity of a
tetrahedral molecule. The model Hamiltonian with nine parameters
and the model transition operator with five parameters
provide quite good fits to the published experimental data
of silicon tetrafluoride with standard deviations 1.188 cm$^{-1}$
and 1.775, respectively. This is based on our new methods for 
constructing symmetrized bases [16] and for 
removing both the spurious states in the wavefunction space and the 
spurious components in the Hamiltonian [23]. For comparison, we also
studied this molecule in the boson-realization model, and obtained
the corresponding standard deviations 1.985 cm$^{-1}$ and 1.745
[28]. Through this example of applications of the model, we
believe that our method for treating spurious states is useful for 
the model for other polyatomic molecules. 

It is shown that anharmonic resonances, such as Darling-Dennison and Fermi
resonances, are very important in descriptions of highly excited 
vibrational states in molecules. Darling-Dennison resonances can
be included in the model by adding higher-order terms of the operators 
of Eq.(2.1). Fermi resonances can be taken into account using perturbation
theory. It should be pointed out that Fermi resonances can be easily 
included in the extended local mode model [30,31] and the boson-realization 
model [17], while they are described by the nondiagonal matrix elements 
of Majorana operators in U(4) algebraic model. The coupling parameters in 
Hamiltonian we have obtained can be related to force field constants to be 
used in conjunction with a kinetic energy operator in a Schr\"odinger equation
[32]. We will consider the physical meaning of the parameters in
future investigations.

\acknowledgments 
The authors would like to thank Prof. Jin-Quan Chen and 
Dr. Jia-Lun Ping for useful discussions. This work was supported by 
the National Natural Science Foundation of China and Grant No. 
LWTZ-1298 of the Chinese Academy of Sciences.

\vspace{30mm}
\begin{center}
{\small Table I. Observed and calculated energy levels and 
relative intensity }

\vspace{3mm}
\begin{tabular}{ccc|ccc}
\hline
\hline
   & Obs.[25] &    &   & Calc. &     \\  \hline
$~\Gamma~$& $~E$ (cm$^{-1}~$)
&Intensity&$~V~$&$~E$ (cm$^{-1}~$)&Intensity  \\ \hline
$E$    & 264.2   &      &1&264.415 &      \\
$F_2$ & 388.4448& 500  &1&388.858 &500.059 \\
$F_2$ & 776.3   & 0.9  &2&775.327 &2$\times10^{-6}$   \\
$A_1$ &800.8    &      &1&799.770 &        \\
$F_2$ &1031.3968& 5000 &1&1029.677&4999.867 \\
$E$   &1064.2   &      &2&1064.186&        \\
$F_2$ &1164.2   & 1.4  &3&1164.169&3$\times10^{-5}$ \\
$F_2$ &1189.7   & 40   &2&1188.631&40.065  \\
$F_2$ &1294.05  & 2.4  &2&1293.903&$4 \times 10^{-4}$ \\
$F_2$ &1418.75  & 0.1  &2&1418.533&0.005  \\
$F_2$ &1804.5   & 0.7  &3&1804.706&5$\times10^{-5}$  \\
$F_2$ &1828.17  & 7    &2&1828.745&3.888   \\
$F_2$ &2059.1   & 1.2  &2&2058.010&3.744   \\
$F_2$ &2602.55  & 0.007&4&2603.789&$4 \times 10^{-8}$ \\
$F_2$ &2623.8   & 0.015&3&2623.678&0.001  \\
$F_2$ &3068.5   & 0.015&3&3069.126&1$\times10^{-5}$   \\
\hline
\hline
\end{tabular}
\end{center}


\begin{references}


\bibitem{1} F. Iachello, Chem. Phys. Lett. {\bf 78}(1981), 581.
\bibitem{2} F. Iachello and R. D. Levine, J. Chem. Phys. {\bf 77} (1982), 3046.
\bibitem{3} F. Iachello and R. D. Levine, Algebraic Theory of Molecules,
Oxford Uni., Oxford, 1995.
\bibitem{4} S. Oss, Adv. Chem. Phys., {\bf 93} (1996), 455.
\bibitem{5} O. S. van Roosmalen, F. Iachello, R. D. Levine, and 
A. E. L. Dieperink, J. Chem. Phys. {\bf 79} (1983), 2515.
\bibitem{6} O. S. van Roosmalen, I. Benjamin, 
and R. D. Levine, J. Chem. Phys. {\bf 81} (1984), 5986.
\bibitem{7} F. Iachello and S. Oss, Phys. Rev. Lett. {\bf 66} (1991), 2976.
\bibitem{8} J. L. Ping and J. Q. Chen, Ann. Phys. (N.Y.) {\bf 255} (1997), 75.
\bibitem{9} T. Sako and K. Yamanouchi, Chem. Phys. Lett. {\bf 264} (1997), 403.
\bibitem{10} I. L. Cooper and R. K. Gupta, Phys. Rev. A {\bf 55} (1997), 4112.
\bibitem{11} F. Iachello and S. Oss, J. Chem. Phys. {\bf 99} (1993), 7337.
\bibitem{12} F. Iachello and S. Oss, J. Chem. Phys. {\bf 104} (1996), 6956.
\bibitem{13} A. Frank, R. Lemus, R. Bijker, F. P\'erez-Bernal, and J. M. Arias,
Ann. Phys. (N.Y.) {\bf 252} (1996), 211.
\bibitem{14} F. P\'erez-Bernal, R. Bijker, A. Frank, R. Lemus, and J. M. Arias, 
Chem. Phys. Lett. {\bf 258} (1996), 301.
\bibitem{15} M. S. Child and L. Halonen, Adv. Chem. Phys. {\bf 57} (1984), 1.
\bibitem{16} Z. Q. Ma, X. W. Hou, and M. Xie, Phys. Rev. A {\bf 53} (1996),
2173.
\bibitem{17} X. W. Hou, M. Xie, and Z. Q. Ma, Phys. Rev. A, {\bf 55} (1997), 
3401.
\bibitem{18} X. W. Hou, M. Xie, and Z. Q. Ma, 
 Inter. J. Theor. Phys. {\bf 36} (1997), 1153.
\bibitem{19} R. Bijker, A. E. L. Dieperink, and A. Leviatan, Phys. Rev. A
{\bf 52} (1995), 2786.
\bibitem{20} C. Leroy and F. Michelot, J. Mol. Spectrosc. {\bf 151} (1992), 71.
\bibitem{21} L. Halonen and M. S. Child, Mol. Phys. {\bf 46} (1982), 239.
\bibitem{22} C. Leroy and V. Boujut,J. Mol. Spectrosc. {\bf 181} (1997), 
127.
\bibitem{23} M. Xie, X. W. Hou, and Z. Q. Ma, Chem. Phys. Lett. 
{\bf 262} (1996), 1.
\bibitem{24} R. Lemus and A. Frank, J. Chem. Phys. {\bf 101} (1994), 8321.
\bibitem{25} R. S. McDowell, M. J. Reisfeld, C. W. Patterson, B. J.
Krohn, M. C. Vasquez, and G. A. Laguna, J. Chem. Phys. {\bf 77} (1982),
4337.
\bibitem{26} J. Q. Chen, A. Klein, and J. L. Ping, J. Math. Phys.
{\bf 37} (1996), 2400.
\bibitem{27} J. Q. Chen and J. L. Ping, J. Math. Phys. {\bf 38} (1997), 
387.
\bibitem{28} X. W. Hou, M. Xie, S. H. Dong, and Z. Q. Ma, 
Ann. Phys. (N.Y.) (1998), in press.
\bibitem{29} J. Q. Chen, F. Iachello, and J. L. Ping, J. Chem. Phys.
{\bf 104} (1996), 815.
\bibitem{30} T. Lukka, E. Kauppi, and L. Halonen, J. Chem. Phys.
{\bf 102} (1995), 5200.
\bibitem{31} L. Halonen, J. Chem. Phys. {\bf 106} (1997), 7931.
\bibitem{32} L. Halonen, J. Chem. Phys. {\bf 106} (1997), 831.
\end{references}
\end{document}